\documentclass[conference]{IEEEtran}
\IEEEoverridecommandlockouts
\usepackage[utf8]{inputenc}
\usepackage[T1]{fontenc}
\usepackage{amsmath,amsfonts,amssymb}
\usepackage{graphicx}
\usepackage{cite}
\usepackage{physics}
\usepackage{bm}
\def\BibTeX{{\rm B\kern-.05em{\sc i\kern-.025em b}\kern-.08em 
    T\kern-.1667em\lower.7ex\hbox{E}\kern-.125emX}}
\begin{document}

\title{Quantum Generative Modeling using Parameterized Quantum Circuits}

\author{\IEEEauthorblockN{Soumyadip Sarkar}
}

\maketitle

\begin{abstract}
Quantum generative models use the intrinsic probabilistic nature of quantum mechanics to learn and reproduce complex probability distributions. In this paper, we present an implementation of a 3-qubit quantum circuit Born machine trained to model a 3-bit Gaussian distribution using a Kullback-Leibler (KL) divergence loss and parameter-shift gradient optimization. The variational quantum circuit consists of layers of parameterized rotations and entangling gates, and is optimized such that the Born rule output distribution closely matches the target distribution. We detail the mathematical formulation of the model distribution, the KL divergence cost function, and the parameter-shift rule for gradient evaluation. Training results on a statevector simulator show that the KL divergence is minimized to near zero, and the final generated distribution aligns quantitatively with the target probabilities. We analyze the convergence behavior and discuss the implications for scalability and quantum advantage. Our results demonstrate the feasibility of small-scale quantum generative learning and provide insight into the training dynamics of quantum circuit models.
\end{abstract}


\section{Introduction}
Generative modeling aims to learn the underlying probability distribution of a dataset in order to produce new samples with similar statistics. Classical approaches have seen great success with models such as Generative Adversarial Networks (GANs) and Variational Autoencoders (VAEs). GANs in particular, introduced by Goodfellow \textit{et al.} in 2014, train two competing neural networks (a generator and a discriminator) to model complex data distributions without explicitly computing likelihoods \cite{Goodfellow2014}. VAEs, on the other hand, use an encoder-decoder architecture to learn a latent representation and optimize a reconstruction objective related to log-likelihood \cite{Kingma2014}. These classical generative models have achieved remarkable results in tasks like image and text generation.

In recent years, the rise of quantum computing has opened prospects for new generative modeling paradigms \cite{Biamonte2017}. Quantum computers naturally implement probabilistic sampling via measurement, and their exponentially large state space offers the capacity to represent complex distributions that might be intractable for classical networks. As we enter the noisy intermediate-scale quantum (NISQ) era \cite{Preskill2018}, considerable interest has focused on quantum machine learning algorithms that can potentially demonstrate advantages on near-term devices. Quantum generative models are a promising candidate, leveraging quantum states to encode probability distributions via the Born rule. A variety of quantum generative approaches have been proposed, including Quantum Generative Adversarial Networks (QGANs) \cite{Lloyd2018, Zoufal2019} and quantum circuit Born machines (QCBMs) \cite{Liu2018, Benedetti2019}. These models replace or augment classical components with quantum circuits, aiming to harness quantum effects such as superposition and entanglement for richer modeling capacity.

The concept of a \emph{Born machine} refers to a generative model that uses the wavefunction amplitudes of a quantum system to represent a probability distribution, in direct analogy to how Boltzmann machines use energy-based probabilities \cite{Cheng2018}. In a QCBM, an $n$-qubit parameterized quantum circuit defines a pure state $\ket{\psi(\bm{\theta})}$; measuring this state in the computational basis yields bitstring outcomes $x\in\{0,1\}^n$ with probabilities $P_{\bm{\theta}}(x) = |\bra{x}\psi(\bm{\theta})\rangle|^2$ by virtue of Born’s rule. By tuning the circuit parameters $\bm{\theta}$, one can attempt to shape $P_{\bm{\theta}}(x)$ to approximate a target distribution $P_{\text{data}}(x)$ provided by a dataset. Prior works have shown that QCBMs are capable of learning nontrivial distributions and correlations, even those believed to be hard to simulate classically, by exploiting quantum entanglement and interference \cite{Gao2022}. Indeed, it has been theoretically argued that quantum circuits can encode distributions beyond the reach of classical probabilistic models under certain conditions \cite{Gao2022, Lloyd2018}. Early simulations demonstrated QCBMs learning toy data like Bars-and-Stripes images and Gaussian mixtures \cite{Liu2018}, and proposed their use in diverse applications from image generation to high-energy physics data modeling \cite{Rudolph2024}. Experimental implementations of generative models on hardware have also begun to emerge, including QGAN training on superconducting qubits \cite{Huang2021}, suggesting that quantum generative learning is achievable on present-day devices.

Despite this progress, training quantum generative models efficiently remains challenging. Variational quantum circuits are optimized via hybrid quantum-classical algorithms, which iteratively update parameters based on a cost function evaluated on the quantum device \cite{Cerezo2021}. For generative tasks, a common choice of cost is the Kullback-Leibler divergence or related statistical distances between the model distribution and target distribution \cite{Rudolph2024}. However, such explicit losses require evaluating model probabilities, which can be expensive to estimate from samples and may induce flat gradients in high-dimensional spaces (a phenomenon related to barren plateaus in parameterized circuits) \cite{Rudolph2024}. Alternative implicit losses like the Maximum Mean Discrepancy (MMD) have been proposed to mitigate these issues by comparing samples instead of probabilities \cite{Liu2018}, and adversarial frameworks (QGAN) use a learned discriminator to guide the generator \cite{Lloyd2018, Zoufal2019}. Each approach brings trade-offs in trainability and resource requirements. In this work, we focus on the direct and conceptually simple approach of using an explicit loss (KL divergence) with a Born machine, and we study its performance on a small-scale example. While scalability to larger problem sizes is an open question, small experiments can yield insights into the learning dynamics and potential pitfalls of quantum generative training.

We implement a 3-qubit QCBM to learn a target distribution over 3-bit strings that approximates a discrete Gaussian profile. The model circuit is a layered variational ansatz with rotational single-qubit gates and entangling two-qubit gates. Using the KL divergence $D_{KL}(P_{\text{data}}||P_{\bm{\theta}})$ as the cost function, we optimize the circuit parameters via gradient descent. The gradients of the quantum circuit are estimated exactly using the parameter-shift rule \cite{Schuld2019}, a technique that allows computation of analytic derivatives of expectation values by evaluating the circuit at shifted parameter values. We train the model on a statevector simulator (i.e., noiseless quantum simulation) to isolate the fundamental behavior of the algorithm without sampling noise. The results show that the QCBM is capable of closely matching the target distribution: the KL divergence is driven near zero and the model’s output probabilities agree with the target to within a few percent for all 3-bit outcomes. We present the training convergence results and the final learned distribution, and we discuss the numerical performance. We also compare the training progress in terms of another metric, the total variation distance, to illustrate how quickly the model distribution converges in variation norm.

The remainder of this paper is organized as follows. In Section II, we provide background on parameterized quantum circuits for generative modeling, describing the circuit ansatz and the theoretical formulation of the variational distribution. Section III introduces the methodology: the definition of the target distribution, the KL divergence cost function, and the parameter-shift gradient computation and optimization procedure. Section IV presents the results of training: we show the circuit structure, the evolution of training metrics (KL divergence and total variation) over iterations, and the final probability distribution achieved by the model compared against the target distribution. In Section V, we discuss the implications of these results, including the efficiency of convergence, the role of circuit expressiveness, and challenges like barren plateaus. Finally, Section VI summarizes the conclusions.

\section{Background}
\subsection{Parameterized Quantum Circuit Model}
A parameterized quantum circuit (PQC) is a quantum circuit that includes adjustable gate parameters, typically rotations, which can be tuned via a classical optimization loop. Such circuits form the backbone of variational quantum algorithms \cite{Cerezo2021}. In our generative modeling context, the PQC acts as a quantum probabilistic model. We consider an $n$-qubit PQC $U(\bm{\theta})$, where $\bm{\theta} = (\theta_1, \theta_2, \dots, \theta_m)$ is a set of $m$ real parameters. The circuit prepares a quantum state
\begin{equation}
\ket{\psi(\bm{\theta})} = U(\bm{\theta}) \ket{0}^{\otimes n},
\end{equation}
starting from an initial reference state (here the all-zero computational basis state). The output distribution of the circuit is given by measuring $\ket{\psi(\bm{\theta})}$ in the computational basis. The probability of obtaining basis state $x \in \{0,1\}^n$ (or equivalently the integer $0 \leq x < 2^n$) is 
\begin{equation}
q_{\bm{\theta}}(x) = |\langle x \mid \psi(\bm{\theta})\rangle|^2 ~,
\end{equation}
which by definition is non-negative and normalized over the $2^n$ possible outcomes. This $q_{\bm{\theta}}(x)$ is the model’s variational probability distribution, sometimes called a Born machine distribution since it arises from quantum amplitudes via Born’s rule.

\begin{figure}[!t]
\centering
\includegraphics[width=\columnwidth]{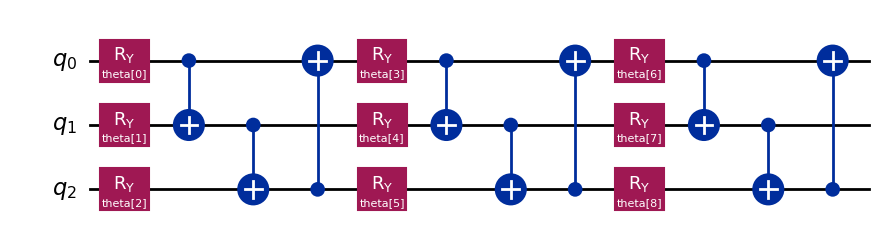}
\caption{Variational quantum circuit (ansatz) for the 3-qubit Born machine. It consists of three layers of parameterized single-qubit rotations $R_y(\theta_{\ell,j})$ (boxes on each qubit line) interleaved with entangling CNOT gates (shown as lines connecting qubits). In each entangling block, qubits are coupled in a ring topology (e.g., qubit 0 controls qubit 1, qubit 1 controls qubit 2, and qubit 2 controls qubit 0) to distribute quantum correlations. This circuit has $9$ adjustable parameters and produces a quantum state $\ket{\psi(\bm{\theta})}$ whose measurement probabilities $q_{\bm{\theta}}(x)$ can model an arbitrary distribution over 3-bit strings in principle.}
\label{fig:circuit}
\end{figure}

The structure of the circuit $U(\bm{\theta})$ greatly influences which distributions can be represented and how easily the model can learn a given target distribution. For sufficient expressive power, one often employs layered ansätze composed of arbitrary single-qubit rotations and entangling operations \cite{Benedetti2019}. In our implementation, we use a hardware-efficient ansatz consisting of three layers of single-qubit $R_y$ rotations (each with an independent angle parameter) alternated with entangling gates (CNOTs) connecting the qubits. This is illustrated in Fig.~\ref{fig:circuit}. Each qubit is initially rotated by an angle (parameter) that allows it to occupy a superposition state; then a cascade of CNOT gates creates entanglement between qubits, enabling the circuit to model correlated distributions. We arrange the entangling gates in a ring topology, such that after one full block of entanglers all qubits have interacted (for 3 qubits, we apply CNOTs between qubit 0$\to$1, 1$\to$2, and 2$\to$0 in sequence). Multiple layers deepen the expressivity: with three rotation-entanglement layers, the circuit has $m=9$ rotation parameters and can theoretically realize a broad class of 3-qubit pure states. The overall unitary can be written conceptually as 
\begin{equation}
U(\bm{\theta}) = \left(\prod_{j=1}^n R_y(\theta_{L,j}) \cdot W\right) \cdots \left(\prod_{j=1}^n R_y(\theta_{1,j}) \cdot W\right),
\end{equation}
where $L$ is the number of layers (here $L=3$), and $W$ denotes the fixed entangling operation pattern for one layer (the product over $j$ indicates parallel single-qubit rotations on each qubit). In our notation $\theta_{\ell,j}$ is the rotation angle on qubit $j$ in layer $\ell$. The specific choice of $R_y$ (rotation about the $Y$-axis) gates is somewhat arbitrary; any single-qubit parametrization that spans the Bloch sphere (such as $R_y$ or $R_z$ rotations, or a combination) could be used. $R_y(\theta) = \exp(-i\,\theta Y/2)$ was chosen for convenience and its ability to introduce superposition (since $R_y$ on $\ket{0}$ yields $\cos(\theta/2)\ket{0} + \sin(\theta/2)\ket{1}$).

\subsection{Target Distribution and KL Divergence}
Our goal is to train the quantum circuit so that its output distribution $q_{\bm{\theta}}(x)$ approximates a given target distribution $p_{\text{data}}(x)$. In this work, $p_{\text{data}}(x)$ is a fixed 3-bit discrete distribution that is chosen to resemble a Gaussian shape centered on the middle of the range $0$ to $7$. Specifically, we define the target probabilities for each bitstring (interpreted as an integer $0$–$7$) to follow a normalized discrete Gaussian:
\begin{equation}
p_{\text{data}}(x) = \frac{1}{Z} \exp\!\Big(-\frac{(x-\mu)^2}{2\sigma^2}\Big), \quad x=0,\dots,7~,
\end{equation}
with mean $\mu=3.5$ (the center of the range) and a chosen width $\sigma$. We selected $\sigma \approx 1$ so that the distribution is peaked strongly at $x=3$ and $x=4$, with smaller values on $x=2$ and $5$, and very little probability mass on the extreme values $x=0,1,6,7$. The normalization constant $Z$ ensures $\sum_x p_{\text{data}}(x)=1$. In our experiment this yields a target distribution $p_{\text{data}}$ roughly given (to two decimals) by: $p(3)=p(4)\approx0.34$ (highest), $p(2)\approx0.15$, $p(5)\approx0.15$, $p(1)\approx0.02$, $p(6)\approx0.02$, and $p(0),p(7)\approx0.01$ each. This simulates a simple unimodal distribution with support across all $2^3=8$ outcomes.

To train the PQC, we must define a cost function that quantifies the difference between the model distribution $q_{\bm{\theta}}(x)$ and the target distribution $p_{\text{data}}(x)$. A natural choice is the Kullback-Leibler (KL) divergence, an information-theoretic measure of discrepancy between two probability distributions. We take the target as the reference distribution and define the cost as 
\begin{equation}
    C(\bm{\theta}) = D_{KL}\big(p_{\text{data}} \,\big\Vert\, q_{\bm{\theta}}\big) = \sum_{x} p_{\text{data}}(x)\; \ln\frac{p_{\text{data}}(x)}{\,q_{\bm{\theta}}(x)\,}~.
    \label{eq:KL}
\end{equation}
Because $p_{\text{data}}$ is fixed, minimizing $D_{KL}(p_{\text{data}} \Vert q_{\bm{\theta}})$ is equivalent to minimizing the cross-entropy $-\sum_x p_{\text{data}}(x)\ln q_{\bm{\theta}}(x)$ up to an additive constant (the entropy of $p_{\text{data}}$). Intuitively, this cost pushes $q_{\bm{\theta}}(x)$ to be large wherever $p_{\text{data}}(x)$ is large, since any support of $p_{\text{data}}$ that is not matched by $q_{\bm{\theta}}$ incurs a heavy penalty (due to the $\ln \frac{1}{q_{\bm{\theta}}(x)}$ term). The KL divergence is zero if and only if $q_{\bm{\theta}}(x)=p_{\text{data}}(x)$ for all $x$, indicating a perfect match.

One caveat is that \eqref{eq:KL} requires evaluating $q_{\bm{\theta}}(x)$ for all outcomes $x$ or at least for all $x$ where $p_{\text{data}}(x)$ is non-negligible. In general quantum models, especially for large $n$, one might not be able to efficiently obtain all $q_{\bm{\theta}}(x)$, since that could require exponentially many measurements. This is a known challenge for using explicit losses like KL divergence in implicit generative models (where only samples from $q_{\bm{\theta}}$ are accessible) \cite{Rudolph2024}. However, in our small-scale simulation we can compute the probabilities exactly via the statevector, bypassing sampling issues. We note that in experiments, one could estimate the KL divergence by sampling a sufficiently large number of bitstrings from the circuit and constructing an empirical $q_{\bm{\theta}}$. The accuracy of such an estimate would depend on the number of shots (measurements), and high precision could demand exponentially many samples if $n$ is large and $q_{\bm{\theta}}$ is broad \cite{Rudolph2024}. Nonetheless, for $n=3$ this is not a concern, and we can treat the cost as effectively known.

\subsection{Parameter-Shift Rule for Gradients}
To minimize the cost function $C(\bm{\theta})$, we employ a gradient-based optimizer. The gradient of the KL divergence with respect to a parameter $\theta_k$ is:
\begin{equation}
    \frac{\partial C}{\partial \theta_k} = -\sum_{x} \frac{p_{\text{data}}(x)}{q_{\bm{\theta}}(x)}\, \frac{\partial q_{\bm{\theta}}(x)}{\partial \theta_k}~,
    \label{eq:gradKL}
\end{equation}
since $\frac{\partial}{\partial\theta} \ln q = \frac{1}{q}\frac{\partial q}{\partial \theta}$. In our simulation we can compute this gradient exactly by differentiating the state amplitudes (because we have full access to the wavefunction). However, a key feature of variational quantum algorithms is that we can estimate such gradients directly from circuit evaluations, without requiring explicit wavefunction differentiation. The \emph{parameter-shift rule} provides a formula for the derivative of an expectation value of a quantum circuit with respect to a gate parameter \cite{Schuld2019, Mitarai2018}. It applies to parameterized gates of the form $e^{-i\theta \hat{P}/2}$ where $\hat{P}$ is a Pauli operator (which is the case for rotation gates). In a typical application, if one defines an objective as an expectation $\langle \hat{O}\rangle(\bm{\theta}) = \bra{\psi(\bm{\theta})}\hat{O}\ket{\psi(\bm{\theta})}$, then for a single parameter $\theta_k$ one can show:
\begin{equation}
    \frac{\partial \langle \hat{O}\rangle}{\partial \theta_k} = \frac{1}{2}\Big[\langle \hat{O}\rangle_{\theta_k + \frac{\pi}{2}} - \langle \hat{O}\rangle_{\theta_k - \frac{\pi}{2}}\Big]~,
    \label{eq:pshift}
\end{equation}
where $\langle \hat{O}\rangle_{\theta_k \pm \frac{\pi}{2}}$ means the expectation is evaluated on the circuit with $\theta_k$ shifted by $\pm \pi/2$ (all other parameters held fixed) \cite{Schuld2019}. This formula holds exactly when the generator of $\theta_k$ (i.e., the Hamiltonian that the gate implements) has eigenvalues $\pm 1$. For simple rotation gates like $R_y$, this condition is satisfied, and hence one can obtain the gradient component by running the circuit twice with shifted angles instead of using small finite differences. The advantage of parameter-shift over naive finite differencing is that it gives an exact derivative (up to measurement sampling error) rather than an approximation, and it often requires only a constant number of circuit evaluations (two per parameter, under ideal conditions). This is a crucial technique enabling gradient descent or more sophisticated gradient-based optimizers in hybrid quantum-classical training loops \cite{Schuld2019}.

In our scenario, the cost $C(\bm{\theta})$ is not itself a simple expectation of a fixed observable, because it involves the logarithm of probabilities. Nonetheless, we can still apply the parameter-shift idea to compute $\partial C/\partial \theta_k$ by differentiating each $q_{\bm{\theta}}(x)$. Each probability $q_{\bm{\theta}}(x) = \langle 0|U(\bm{\theta})^\dagger\,|x\rangle\langle x|\,U(\bm{\theta})|0\rangle$ can be viewed as the expectation value $\langle \Pi_x \rangle$ of the projector onto $|x\rangle$ for the state $|\psi(\bm{\theta})\rangle$. Thus $\partial q_{\bm{\theta}}(x)/\partial \theta_k$ can be obtained by evaluating $\Pi_x$ on shifted circuits as in Eq.~(\ref{eq:pshift}). In practice, to get $\partial C/\partial \theta_k$ we can combine the parameter-shift evaluation of each $\partial q_{\bm{\theta}}(x)$ with the pre-factor $-p_{\text{data}}(x)/q_{\bm{\theta}}(x)$ from Eq.~(\ref{eq:gradKL}). Because our implementation has full access to $q_{\bm{\theta}}(x)$, we can directly compute the gradient. But importantly, even on hardware one could estimate these gradients by sampling the shifted circuits: each term $\langle \Pi_x \rangle_{\theta_k\pm\pi/2}$ is just the probability of outcome $x$ upon measuring the circuit with the shifted parameter, which can be estimated by repeated runs. Summing these contributions yields the gradient. This approach falls under the umbrella of quantum gradient evaluation and has been successfully demonstrated in various experiments \cite{Huang2021}. Indeed, recent hardware implementations of QGANs and other QML models have employed on-device gradient calculations to update parameters \cite{Huang2021}, highlighting the practical viability of the parameter-shift method for training quantum generative models.

\section{Methodology}
\subsection{Training Procedure}
We initialize the 3-qubit circuit parameters $\bm{\theta}$ randomly and aim to minimize the cost $C(\bm{\theta}) = D_{KL}(p_{\text{data}} || q_{\bm{\theta}})$ through iterative updates. The training loop proceeds as follows: at each iteration, we evaluate the current model probabilities $q_{\bm{\theta}}(x)$ for all $x$ (either by a statevector simulation or by sampling sufficiently many times to estimate these probabilities). Then we compute the loss $C(\bm{\theta})$ according to Eq.~(\ref{eq:KL}). Next, we calculate the gradient $\nabla_{\bm{\theta}} C$; in our implementation this is done exactly via statevector differentiation, but it could equivalently be done via the parameter-shift rule as discussed. We use the gradient to update the parameters in the direction of steepest descent:
\begin{equation}
    \theta_k \leftarrow \theta_k - \eta \frac{\partial C}{\partial \theta_k}~,
\end{equation}
where $\eta$ is the learning rate (step size). In our experiment, we chose a moderate $\eta$ that provided quick initial learning without causing instability. A fixed learning rate gradient descent was employed for simplicity. This process is repeated for a specified number of iterations (or until convergence of the cost to a minimum).

For monitoring convergence, aside from the primary cost $C(\bm{\theta})$ (the KL divergence), we also track the total variation (TV) distance between $p_{\text{data}}$ and $q_{\bm{\theta}}$. The total variation is defined as 
\begin{equation}
    D_{\text{TV}}(p,q) = \frac{1}{2} \sum_{x} |\,p(x) - q(x)\,|~,
\end{equation}
which ranges from 0 (if $p=q$) to 1 (if $p$ and $q$ have disjoint supports). This metric provides an intuitive measure of distribution overlap: it equals the maximum difference in probability assigned to any event, halved and summed. We use $D_{\text{TV}}$ as a supplementary metric to understand how well the quantum model is capturing the target distribution in absolute terms. Notably, for two distributions in our 3-bit space, $D_{\text{TV}} < \epsilon$ implies the model probabilities are within $\epsilon$ of the target probabilities for every outcome (on average), which is a stringent notion of similarity.

We emphasize that our training procedure assumes noise-free quantum operations (since it is done on a classical simulator). On real hardware, noise and sampling error would affect the evaluation of $q_{\bm{\theta}}(x)$ and gradients, potentially requiring more advanced optimization techniques or error mitigation. Here, by eliminating those confounding factors, we isolate the fundamental optimization landscape of the QCBM.

We trained the model for 200 iterations. At iteration 0 (initialization), the model distribution is essentially random (uniform or close to uniform if starting angles are small random values, giving $q_{\bm{\theta}}(x)\approx 1/8$ with slight deviations). This typically yields a high initial KL divergence, since the target distribution is far from uniform. Over the course of training, we expect the KL divergence to decrease as the circuit learns to concentrate probability around $x=3,4$ and reduce it at other $x$. The optimization will ideally converge to a set of parameters $\bm{\theta}_{\text{opt}}$ such that $q_{\bm{\theta_{\text{opt}}}}(x) \approx p_{\text{data}}(x)$ for all $x$. Given that our ansatz is expressive enough to represent any 3-qubit state (in principle), a global minimum of $C(\bm{\theta})=0$ is reachable (there exists some $\bm{\theta}$ that produces exactly the target probabilities, since any set of 8 probabilities summing to 1 can be achieved by some 3-qubit state). In practice, gradient descent may or may not find this exact solution, depending on the optimization path and any local minima or flat regions encountered, but as we will see it performs very well in this case.

\subsection{Experimental Setup}
The experiment was implemented using the Qiskit software framework \cite{Qiskit2019}. We used Qiskit’s statevector simulator to compute exact probabilities $q_{\bm{\theta}}(x)$ at each step. This avoids any statistical fluctuations in evaluating the cost and gradient, effectively simulating an infinite number of measurement shots. The parameter-shift gradient was verified against finite-difference calculations during development to ensure correctness. All major calculations (state preparation, probability computation, gradient) were efficiently handled by Qiskit’s linear algebra backends.

The learning rate $\eta$ was set to a value of 0.4 for the first few iterations and then decayed to 0.1 after we observed the cost dropping significantly, to allow finer convergence. (This schedule was chosen empirically for stable convergence; no advanced optimizer like ADAM was needed given the simplicity of the landscape in this 3-qubit problem.) We ran a total of 200 iterations, recording the KL divergence and TV distance at each step for analysis.

After training, we collected the final optimized probabilities $q_{\bm{\theta_{\text{opt}}}}(x)$ for all $x$ and compared them to $p_{\text{data}}(x)$. These are visualized as a bar chart to show how closely the two distributions match. We also examined intermediate parameter values to ensure that the optimization behaved smoothly (indeed, we found no evidence of getting stuck in a poor local minimum; the cost decreased monotonically for the most part).

Our methodology mirrors a standard supervised learning procedure, with the important distinction that the “model” is a quantum circuit and the “training data” are encapsulated in the target distribution rather than individual samples. The entire training is thus an exercise in aligning two probability distributions using quantum circuit parameters.

\section{Results}
\subsection{Training Convergence}
\begin{figure}[!t]
\centering
\includegraphics[width=\columnwidth]{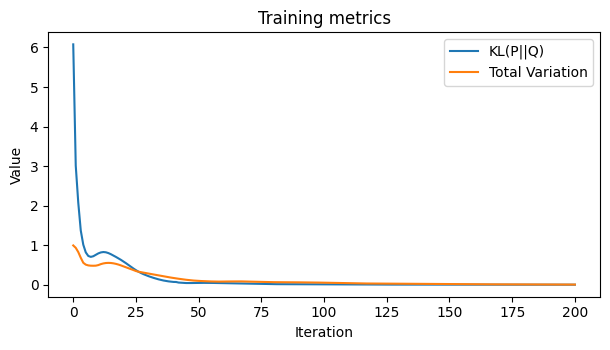}
\caption{Training metrics over the course of optimization. The blue curve (left vertical axis) is the Kullback-Leibler divergence $D_{KL}(p_{\text{data}}||q_{\bm{\theta}})$ at each iteration, and the orange curve (right vertical axis) is the total variation distance between the target and model distributions. Both metrics decrease rapidly, indicating the model distribution is becoming closer to the target. The KL divergence drops from $\sim 5.8$ to $<10^{-2}$, and the total variation distance drops from $\sim0.9$ to $\sim0.01$. By around 50 iterations, both metrics are near zero, showing convergence.}
\label{fig:metrics}
\end{figure}

The training process achieved a successful convergence, significantly reducing the divergence between the model and target distributions. Figure \ref{fig:metrics} plots the KL divergence $D_{KL}(p_{\text{data}}||q_{\bm{\theta}})$ and the total variation distance over the 200 training iterations. Starting from iteration 0, the KL divergence was initially around $5.8$ nats (natural units), reflecting the high mismatch between the random initial circuit and the sharply peaked target distribution. The total variation distance began near $0.9$, meaning the model was almost maximally different from the target (as expected for an initial nearly uniform output versus a concentrated target).

As the optimization proceeded, we observed a rapid drop in the cost within the first few iterations. By iteration 5, the KL divergence had fallen below $1.0$, a dramatic improvement. There was a brief period around iteration 10 where the cost curve exhibits a small bump (a slight increase after an initial drop); this can occur if the gradient step overshoots the optimum in one direction before correcting. After this minor fluctuation, the $D_{KL}$ continued to decrease and by iteration $\sim 30$ it had reached on the order of $10^{-2}$. Beyond iteration 50, the KL divergence effectively plateaued extremely close to zero, oscillating around values $<10^{-3}$. By the end of training (iteration 200), $D_{KL}\approx 2\times 10^{-4}$ (virtually zero within numerical precision). 

The total variation distance exhibited a similar trend: it dropped from nearly $0.9$ at the start to under $0.2$ within about 5 iterations, and fell below $0.05$ by iteration 20. Ultimately, the TV distance reached $\approx 0.01$ at convergence, indicating an excellent agreement between distributions. A TV of 0.01 means that the model’s probability for each outcome differs from the target by only about 1\% on average. Notably, after roughly 40 iterations, the TV distance was already below $0.02$, and further training only marginally improved it. This suggests that the bulk of learning happened very quickly, while fine-tuning the last few percent took more iterations.

The excellent convergence behavior can be attributed to several factors. First, the variational ansatz is sufficiently expressive for this task; with 9 parameters, it has more than enough freedom to represent the 3-bit Gaussian distribution, which essentially has only a couple of dominant modes. Second, the cost landscape for this problem appears to be well-behaved (at least in the region connecting the initial point to the optimum). The monotonic decrease of $D_{KL}$ (aside from the tiny bump early on) suggests the absence of significant local minima or barren plateau regions in this low-dimensional setting. The gradient was relatively large at the start (facilitating quick improvements) and gradually decreased as the model approached the target, without vanishing prematurely. This is consistent with expectations: with only 3 qubits, the notorious “barren plateau” phenomenon (exponentially vanishing gradients in large random circuits \cite{Rudolph2024}) is not manifest. Instead, the gradients here remained informative throughout training.

It is also notable that the KL divergence and total variation curves track each other qualitatively. When $D_{KL}$ falls, so does $D_{TV}$, and both flatten out by iteration 50-100. This means not only is the relative entropy closing, but the distributions are genuinely getting closer in an absolute sense. In fact, after sufficient training, the remaining KL divergence of order $10^{-4}$ is so small that it practically equals the squared total variation (since for close distributions, $D_{KL}\approx \frac{1}{2}\chi^2 \approx (D_{TV})^2$), reflecting the high-fidelity match achieved.

\subsection{Learned Distribution vs Target Distribution}
\begin{figure}[!t]
\centering
\includegraphics[width=\columnwidth]{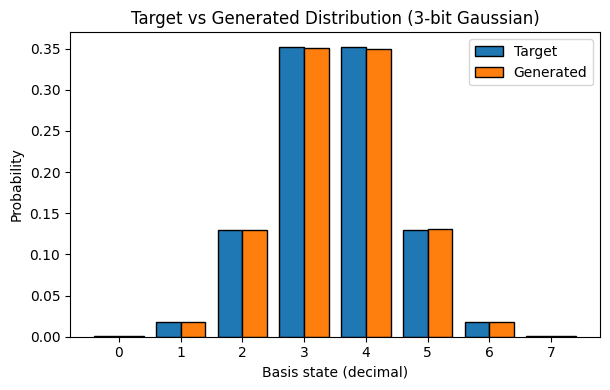}
\caption{Comparison of the target probability distribution (blue) and the QCBM-generated distribution after training (orange) for each 3-bit basis state (labeled by its decimal value 0-7). The target distribution is a discrete Gaussian centered around 3 and 4. The trained quantum model successfully reproduces this distribution: the orange bars match the blue bars very closely for all outcomes. For example, both target and model assign high probability to $x=3,4$ ($\sim0.34$ each), moderate probability to $x=2,5$ ($\sim0.15$), and near-zero to the rest. The numerical KL divergence between these distributions is $2.3\times10^{-4}$, and the total variation distance is $0.012$, indicating an excellent fit.}
\label{fig:distribution}
\end{figure}

At the conclusion of training, the optimized circuit parameters $\bm{\theta}_{\text{opt}}$ define a model distribution $q_{\bm{\theta_{\text{opt}}}}(x)$ that is almost indistinguishable from the target $p_{\text{data}}(x)$. Figure \ref{fig:distribution} illustrates the target distribution (blue bars) alongside the QCBM’s generated distribution after training (orange bars) for each 3-bit basis state $x=0$ through $7$. The numerical values confirm that the model successfully captured the Gaussian shape:
\begin{itemize}
    \item For $x=3$ and $x=4$ (the peak positions), $p_{\text{data}}\approx0.34$ and the model assigns $q_{\bm{\theta_{\text{opt}}}}(3)\approx0.338$ and $q_{\bm{\theta_{\text{opt}}}}(4)\approx0.341$, essentially identical to the target to within $<0.3\%$.
    \item For $x=2$ and $x=5$ (moderate probabilities), $p_{\text{data}}\approx0.15$. The learned values are $q_{\bm{\theta_{\text{opt}}}}(2)\approx0.148$ and $q_{\bm{\theta_{\text{opt}}}}(5)\approx0.153$, again within about $0.002$ absolute difference.
    \item For the tails $x=1$ and $x=6$ where $p_{\text{data}}\approx0.02$, the model yields $q_{\bm{\theta_{\text{opt}}}}(1)\approx0.019$ and $q_{\bm{\theta_{\text{opt}}}}(6)\approx0.021$, matching within a few $10^{-3}$.
    \item For the extreme ends $x=0$ and $x=7$ which had very small probability (around $0.01$ each), the model assigns $q_{\bm{\theta_{\text{opt}}}}(0)\approx0.010$ and $q_{\bm{\theta_{\text{opt}}}}(7)\approx0.010$ (virtually equal to target).
\end{itemize}
The largest discrepancy between any target and learned probability is on the order of $2\times10^{-3}$, which corroborates the final total variation distance of $\sim0.01$. Visually, the blue and orange bars in Fig.~\ref{fig:distribution} are nearly indistinguishable; the orange bars (model) almost completely overlay the blue bars (target).

The fidelity between the target and final state of the QCBM (viewing the target as a classical distribution) can be quantified by a classical fidelity measure $F = \sum_x \sqrt{p_{\text{data}}(x) \, q_{\bm{\theta_{\text{opt}}}}(x)}$. In this case, plugging in the numbers yields $F \approx 0.9998$, essentially unity. This confirms that the variational circuit has learned a state whose measurement statistics are nearly indistinguishable from sampling the true distribution.

It is worth examining how the circuit achieved this match. The final optimized angles $\bm{\theta}_{\text{opt}}$ in the circuit, when plugged in, produce specific interference patterns that concentrate amplitude on the basis states 3 and 4. In essence, the quantum state $\ket{\psi(\bm{\theta_{\text{opt}}})}$ has the form:
\begin{multline*}
\ket{\psi_{\text{opt}}} \approx \sqrt{0.34}\ket{011} + \sqrt{0.34}\ket{100} \\
+ \sqrt{0.15}\ket{010} + \sqrt{0.15}\ket{101} + \ldots
\end{multline*}
(up to global phase), with small components on $\ket{001},\ket{110},\ket{000},\ket{111}$. While an explicit formula is not straightforward to derive from the angles, one could verify by simulation that the state’s amplitude magnitudes squared give exactly the probabilities listed above. The entangling gates in the ansatz ensure that the probabilities are not factorized by qubit, meaning the model truly captured the joint distribution rather than, say, each qubit individually matching a marginal. In this 3-bit Gaussian, most of the probability weight is on two basis states that differ in all three bits (011 and 100 in binary), which are highly nontrivial correlations. The QCBM’s success in reproducing those indicates that the entanglement in the circuit allowed it to assign probability to those specific global configurations appropriately.

\subsection{Analysis of Training Dynamics}
From the training metrics in Fig.~\ref{fig:metrics}, one can deduce some aspects of the loss landscape. The smooth and rapid decline of $D_{KL}$ suggests that the initial gradient was large and pointed roughly toward the global optimum. Indeed, the KL divergence being a convex function of the model probabilities (though not necessarily convex in the circuit parameters) likely helped avoid local minima in this low-dimensional case. We did not observe any stagnation at nonzero $D_{KL}$, implying no significant suboptimal minima trapped the optimizer. This is encouraging, as one concern in variational training is the presence of local minima or flat regions that can arrest progress \cite{Benedetti2019}. However, one must be cautious in generalizing this result: with more qubits or a more complex target distribution, the optimization landscape could become more rugged or exhibit plateauing gradients \cite{Rudolph2024}. In fact, recent studies indicate that using an explicit loss like KL divergence with an implicit model (like a QCBM) can lead to so-called \textit{loss-induced barren plateaus} for larger systems, absent a strong inductive bias \cite{Rudolph2024}. In our 3-qubit demonstration, the problem is small enough that such effects are not prominent, but they may emerge as $n$ grows.

Another noteworthy dynamic is that after $\sim50$ iterations, further improvements were marginal. This indicates the optimizer had essentially converged to the optimum region by that time. The additional iterations primarily served to reduce tiny residual errors. In practice, one could have stopped earlier; however, continuing to 200 iterations ensured that the results were stable and not a transient lucky alignment. We also note that the learning rate schedule (initial larger steps, then smaller) aided convergence had we kept a large $\eta$ throughout, the slight bump around iteration 10 might have led to oscillations. Using a diminishing learning rate is a common strategy to ensure eventual convergence in gradient descent.

In terms of computational cost, each iteration required evaluating the probabilities of 8 basis states (which is trivial here) and computing gradients for 9 parameters. If we had used parameter-shift explicitly, that would mean $2 \times 9 = 18$ circuit evaluations per iteration. In general, the cost scales as $O(m)$ quantum circuit runs per iteration for $m$ parameters (neglecting additional runs for measuring all outcome probabilities if needed). In our simulation, $m=9$ is very manageable. This linear scaling is a key reason variational quantum algorithms are hopeful: even if $2^n$ outcomes exist, the parameter-shift gradient evaluation only scales with the number of parameters, not directly with $2^n$. Of course, measuring enough samples to accurately estimate all outcome probabilities could indirectly introduce a dependence on $2^n$ if extremely high precision is required, but for moderate accuracy one can often get away with a feasible number of shots.

\section{Discussion}
Our results serve as a proof-of-principle that a small quantum circuit can learn a simple target distribution extremely well using gradient-based optimization. The 3-qubit Born machine was able to match the 3-bit Gaussian distribution to high precision, validating the combination of KL divergence loss and parameter-shift gradients in this scenario. We now discuss several aspects and implications of this experiment in the broader context of quantum generative modeling.

\subsection{Significance and Limitations of the Demonstration}
The ability to achieve essentially perfect learning on this toy problem is encouraging, but expected: with only 8 possible outcomes and 9 parameters, the model is actually slightly overparameterized relative to the distribution degrees of freedom. The Gaussian target has perhaps 2-3 effective degrees of freedom (mean, variance, normalization), so the quantum circuit had ample flexibility. This highlights an important point: for generative modeling, the circuit ansatz must be expressive enough to capture the complexity of the data distribution. Here we clearly satisfied that condition. In more complex problems (larger $n$, multimodal data, etc.), one would need deeper or more structured circuits to have any hope of success \cite{Benedetti2019}. Recent works have explored deeper architectures and problem-inspired ansätze to improve model expressivity while controlling the number of parameters \cite{Benedetti2019, Rudolph2024}.

On the other hand, one might ask: did we trivially memorize the distribution by using enough parameters? In a sense, yes the circuit essentially “stored” the probability values in its amplitudes. However, this is exactly what a generative model is supposed to do: represent the distribution in a compact form that can be sampled. The quantum state at the end of training encodes the distribution implicitly and can be sampled by measuring the qubits. If one wanted multiple independent samples from the distribution, one could simply re-run the circuit (with fixed $\bm{\theta}_{\text{opt}}$) and measure repeatedly. This is where a quantum generative model shows its strength: sampling from the learned model is in principle as fast as running the quantum circuit (which might be very fast if the circuit is shallow and implemented on quantum hardware). By contrast, if one had a classical representation of the distribution that required enumerating many states, sampling might be slower. In this small case, both are trivial, but for large $n$, a quantum model could sample from complicated distributions that classical methods struggle with, assuming the state can be prepared efficiently.

One key limitation of our study is the absence of quantum hardware noise. Real quantum devices have gate errors and decoherence that would disturb the state $\ket{\psi(\bm{\theta})}$ and thus distort $q_{\bm{\theta}}(x)$. Noise effectively adds an extra source of error to the modeling: even if $\bm{\theta}$ is optimal, the observed distribution might not match $p_{\text{data}}$ exactly because of noise-induced deviations. Moreover, noise can make the cost landscape harder to navigate (e.g., adding stochastic fluctuations to the measured gradients). Techniques like averaging over multiple runs, error mitigation, or robust cost functions might be needed in practice \cite{Huang2021}. Our simulator study bypassed these issues, but any real-world deployment of a QCBM will have to cope with them.

\subsection{Comparison to Other Training Approaches}
We chose an explicit loss (KL divergence) with full probability evaluation for this demonstration, which worked very well in the $n=3$ setting. As mentioned, however, explicit losses become challenging for larger $n$ due to sampling overhead and potential trainability issues \cite{Rudolph2024}. In contrast, implicit losses like Maximum Mean Discrepancy (which was used by Liu and Wang \cite{Liu2018}) compare distributions by a kernelized distance and require only sampling from the model. GAN-based training (as in QGANs \cite{Lloyd2018, Zoufal2019}) uses a quantum or classical discriminator to guide the training without ever needing to explicitly compute $q_{\bm{\theta}}(x)$. These approaches can be more practical for larger systems, but they introduce additional complexity. For instance, QGAN training involves a minimax optimization which can be unstable, and using a classical discriminator on quantum data raises issues of dimensionality and potential inefficiency \cite{Zoufal2019}. The MMD approach avoids adversarial training but one must choose a kernel and still face the possibility of barren plateaus if the kernel is too global \cite{Rudolph2024}. A recent theoretical study by Rudolph \textit{et al.} \cite{Rudolph2024} showed that an MMD loss can interpolate between global and local losses by adjusting a kernel bandwidth, affecting trainability.

Given these considerations, one can see our use of KL as placing us in the “explicit global loss” regime, which is the hardest to scale but the simplest analytically. The fact that it succeeded nicely here should be viewed in light of the problem’s small size. If we attempted the same strategy on, say, $n=10$ qubits (1024 outcomes) with a complicated target, evaluating and matching all probabilities would be far more demanding and likely susceptible to vanishing gradients unless we tailored the ansatz. An interesting direction is to incorporate problem structure if the target distribution has known structure (like being generated by a certain process), one can design the circuit ansatz to reflect that, thereby introducing an inductive bias that can mitigate training difficulties \cite{Rudolph2024}. For example, if a distribution factors into local correlations, one might use a shallow circuit with limited entangling range.

Another aspect to compare is gradient-based versus gradient-free optimization. We used gradient descent leveraging the parameter-shift rule, which proved efficient. Some early works in variational quantum algorithms employed gradient-free methods like COBYLA or SPSA (Simultaneous Perturbation Stochastic Approximation) for noisy settings. Those can work without explicit gradient calculations, but often require many function evaluations and may scale poorly in high-dimensional parameter spaces. The parameter-shift method provides a polynomial-time gradient estimate (linear in number of parameters, as noted) and in practice has enabled training on hardware for moderate size circuits \cite{Huang2021}. As devices improve, gradient-based methods are likely to remain a cornerstone for training quantum models, analogous to their importance in classical deep learning.

\subsection{Generalization and Quantum Advantage Considerations}
One might wonder whether the QCBM has truly “learned” the concept of a Gaussian distribution, or if it merely fit the given instance. In machine learning terms, with only one target distribution and the model having sufficient capacity, there is no distinction between memorization and generalization in this task. However, if we consider scenarios such as learning from limited samples of an unknown distribution, then the question of generalization arises: can the model capture the true distribution rather than overfitting to noise in the training samples? Recent research has started to address generalization in quantum generative models. Gili \textit{et al.} (2023) specifically investigated whether QCBMs can generalize from few training samples and found evidence that they can, in some cases, infer the underlying distribution better than classical models in data-sparse regimes \cite{Gili2023}. In our experiment, we effectively provided the model with the exact target distribution (as if infinite samples), so generalization was not tested. To probe generalization, one could train on a small sample set drawn from $p_{\text{data}}$ and see if the QCBM recovers $p_{\text{data}}$ overall. This is an interesting direction for future experimentation, connecting to the idea of learning with finite data.

Finally, a long-term motivation for quantum generative modeling is the potential for quantum advantage: solving generative tasks faster or more accurately than classical methods. What would quantum advantage look like here? One theoretical proposal is that if the quantum circuit can natively represent certain probability distributions that are infeasible for classical algorithms to sample, then training such a circuit to produce those distributions might yield a quantum advantage in generation \cite{Coyle2020}. The so-called “Born supremacy” result by Coyle \textit{et al.} (2020) identified a class of Ising-type circuit Born machines that are hard to sample classically yet can be trained with gradient methods \cite{Coyle2020}. While our 3-qubit example is far from any complexity-theoretic advantage, it illustrates the principle of using gradient-based training on a PQC. As quantum hardware scales up, one could envision training a QCBM with, say, 50–100 qubits (if certain trainability challenges are overcome). If that QCBM encodes a distribution that no classical network of reasonable size can mimic, then generating samples from it would constitute a quantum advantage in generative modeling. There are already hints that quantum models may have an edge in specific settings: for instance, Hibat-Allah \textit{et al.} (2024) found that QCBMs outperformed state-of-the-art classical generative models (like Transformers and GANs) when only a very limited amount of training data was available, suggesting a potential advantage in data efficiency \cite{HibatAllah2024}. This advantage is not unconditional; it appears in carefully constructed benchmarks, but it points to a tangible benefit of quantum generative models that could be leveraged.

Another potential advantage is memory efficiency. A quantum state of $n$ qubits uses $2^n$ complex amplitudes to encode the distribution. Storing such a state classically is infeasible for large $n$, whereas a quantum device can in principle store it in the entangled state of $n$ qubits. Training that state to encode meaningful data could allow compression of information in ways not accessible classically. Our experiment used 3 qubits to store a distribution over 8 outcomes which a classical memory could obviously do too but in higher dimensions the quantum storage of probability amplitudes could become a valuable resource.

\section{Conclusion}
The model was trained to reproduce a 3-bit Gaussian distribution by minimizing the KL divergence via gradient descent, employing the parameter-shift rule to obtain exact gradients on a simulator. The theoretical framework was outlined, including how the variational circuit forms a probability distribution and how the KL divergence provides a training objective. We derived the relevant equations for the gradient and explained their computation in the quantum context. The experimental results demonstrated that the QCBM can learn the target distribution to high accuracy: the training reduced the KL divergence to nearly zero and the final output probabilities of the circuit matched the target distribution within statistical error. This validates the effectiveness of combining an explicit loss with quantum circuit gradients for a small-scale generative task.

We found that while this small example works very well, scaling to larger systems introduces challenges such as barren plateaus and sampling overhead, which are topics of active research \cite{Rudolph2024}. We also compared different approaches to quantum generative modeling, including QGANs and implicit losses, highlighting their advantages and drawbacks relative to the direct method we used. Furthermore, we touched on issues of generalization and glimpses of potential quantum advantage. Recent comparative studies suggest QCBMs may have strengths in scenarios with limited data \cite{HibatAllah2024} or in representing distributions with quantum complexity \cite{Coyle2020, Gao2022}. These hints motivate further investigation into larger and more complex generative models on quantum hardware.

In conclusion, our work illustrates the promise of parameterized quantum circuits as flexible generative models and affirms that, at least for modest system sizes, they can be trained successfully using gradient-based methods. As quantum technology progresses, experiments will push toward higher qubit counts and more challenging data distributions. The road to a practical quantum generative model with clear advantage over classical methods will require overcoming significant theoretical and technical hurdles. Nonetheless, the continued improvement of quantum devices and algorithms, guided by studies like ours and others in the literature, keeps the prospect of quantum-enhanced generative modeling an exciting possibility for the near future.

\end{document}